\documentclass[%
 reprint,
superscriptaddress,
 amsmath,amssymb,
 aps,
]{revtex4-1}

\usepackage{graphicx}
\usepackage{dcolumn}
\usepackage{bm}
\usepackage{braket}
\usepackage[subrefformat=parens]{subcaption}
\usepackage{cancel}
\usepackage{enumerate}
\usepackage{multirow}

\begin{document}

\preprint{APS/123-QED}

\title{A New Method to Extract Information of Near-Threshold Resonances\\
-Uniformized Pole-Sum Representation of Green's Function and T-matrix-}

\author{W. Yamada}
\affiliation{Department of Physics, Faculty of Science, University of Tokyo, 7-3-1 Hongo Bunkyo-ku Tokyo 113-0033, Japan}%
\affiliation{Theory Center, Institute of Particle and Nuclear Studies (IPNS), High Energy Accelerator Research Organization (KEK), 1-1 Oho, Tsukuba, Ibaraki, 205-0801, Japan}%
\author{O. Morimatsu}
\affiliation{Department of Physics, Faculty of Science, University of Tokyo, 7-3-1 Hongo Bunkyo-ku Tokyo 113-0033, Japan}%
\affiliation{Theory Center, Institute of Particle and Nuclear Studies (IPNS), High Energy Accelerator Research Organization (KEK), 1-1 Oho, Tsukuba, Ibaraki, 205-0801, Japan}%
\affiliation{Department of Particle and Nuclear Studies,
Graduate University for Advanced Studies (SOKENDAI),
1-1 Oho, Tsukuba, Ibaraki 305-0801, Japan}%
\date{\today}

\begin{abstract}
  We propose a new, simple model-independent method to extract information of near-threshold resonances, such as complex energies and residues.
  The method is based on the observation that the Green's function and the T-matrix can be represented as the sum of all poles, both bound and resonant poles, in the complex plane of a variable in which the Green's function and the T-matrix are single-valued functions.
  The symmetries of poles, which arise from the unitarity of the S-matrix, naturally impose the sum to obey the proper threshold behaviors.
  The imaginary part of Green's function and the T-matrix are directly related to observables such as scattering cross sections or invariant or missing mass distributions of hadron resonances. Thus we can determine their pole positions and residues by fitting their imaginary part to observables.
  We also test the new method by regarding the imaginary part of the $T$-matrix calculated exactly in a model theory as virtual experimental data.
  As a model theory, we take double-channel meson-baryon scatterings in the chiral unitary model with channels, $\overline{K}N (I=0)$, and $\pi\Sigma (I=0)$.
  By fitting the imaginary part of the $T$-matrix calculated in the model theory by that of the uniformized pole-sum, we obtain the pole positions and residues.
  Comparing the obtained results with those of the exact calculation in the model theory, we conclude that our new method works very well.
\end{abstract}

\maketitle
Resonances and threshold behaviors of hadron scatterings are characteristic non-perturbative phenomena in strong interaction physics. From a mathematical perspective, resonances and hadron scattering processes, threshold behavior, in particular, correspond to poles and branch points of an analytic function, the S-matrix.
Formally, the resonance is defined by the pole of the scattering amplitude, $\mathcal{A}$,
as a Breit-Wigner form \cite{Breit:1936zzb},
\begin{align}
     \mathcal{A}(\sqrt{s}) \sim  \frac{\Gamma_R}{\sqrt{s}-M_R-i\frac{\Gamma_R}{2}},
\end{align}
or as a relativistic Breit-Wigner form  (e.g.~Ref.~\cite{Brown:1992db}),
\begin{align}
     \mathcal{A}(s) \sim \frac{M_R\Gamma_R}{s-M_R^2-iM_R\Gamma_R},
\end{align}
where $s$ is the center-of-mass energy squared, $M_R$ and $\Gamma_R$ are the mass and the width of the resonance, respectively.
These formulas describe observables well if the observed center-of-mass energy, $\sqrt{s}$, is close to the pole mass, $M_R$,
and sufficiently distant from the thresholds.
\par
It is also well known that in the vicinity of the threshold the imaginary part of the scattering amplitude behaves as
  \begin{align}\label{eq:th1}
    &{\rm Im} \mathcal{A}(\sqrt{s})=
    \begin{cases}
      0, &(\sqrt{s}<\varepsilon_1)\\
      ak, &(\sqrt{s}>\varepsilon_1)
    \end{cases}
  \end{align}
at the lowest threshold, and
  \begin{align}\label{eq:th2}
    &{\rm Im} \mathcal{A}(\sqrt{s}))=
    \begin{cases}
      c+\alpha\kappa, &(\sqrt{s}<\varepsilon_i)\\
      c+ ak, &(\sqrt{s}>\varepsilon_i)
    \end{cases}
  \end{align}
at higher thresholds~\cite{Newton}, where $\varepsilon_i$ is the threshold energy, $k$ is the momentum in the center-of-mass frame, $\kappa$ is defined by $k=i\kappa$ and $c$, $a$ and $\alpha$ are real constants.
The interrelationship between the resonances and threshold behaviors create prosperous and sophisticated grounds on hadron physics  (e.g.~Ref.~\cite{Guo}). One typical example is the existence of exotic hadrons (e.g.~Ref.~\cite{Karliner}).
\par
Neither the original Breit-Wigner form nor the relativistic Breit-Wigner form incorporates the proper threshold behaviors, which make it challenging to extract information of near-threshold resonances from experimental data.
Some phenomenological attempts have been made to formulate scattering amplitudes such as Ref.~\cite{Flatte:1976xu}, which modifies the Breit-Wigner form to incorporate both resonance and threshold behaviors.

\begin{gather}
  \mathcal{A}(s)\sim \frac{M_R\sqrt{\Gamma_1\Gamma_2}}{M_R^2-s-iM_R(\Gamma_1+\Gamma_2)},
\end{gather}
where, $\Gamma_i=g_ik_i$, $k_i$ is the momentum in the center-of-mass frame, $g_i$ may be considered as the coupling constant squared for the resonance.

These attempts, however, are far from satisfactory from a theoretical point of view. In this paper we propose a novel approach that naturally and perfectly incorporates both resonance and threshold behaviors in a theoretically sound fashion based on analyticity and unitarity.\\
\par

\begin{figure}[h]
  \centering
  \includegraphics[width=0.5\linewidth]{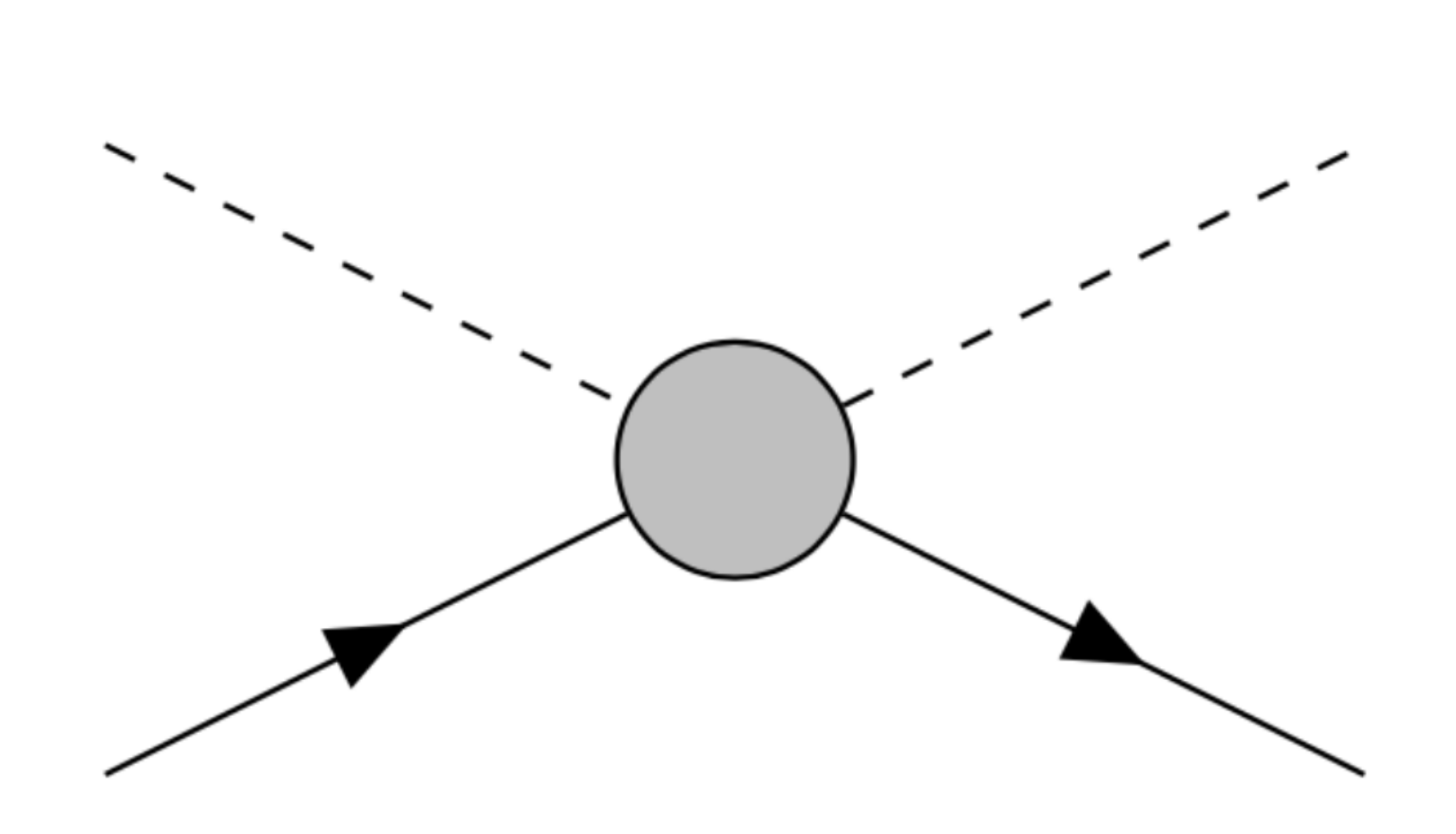}
  \caption{Two-body (meson-baryon) scattering process. the blobe represents the the two-body T-matrix.}
  \label{fig:cross}
\end{figure}

\begin{figure}[h]
  \centering
  \includegraphics[width=0.5\linewidth]{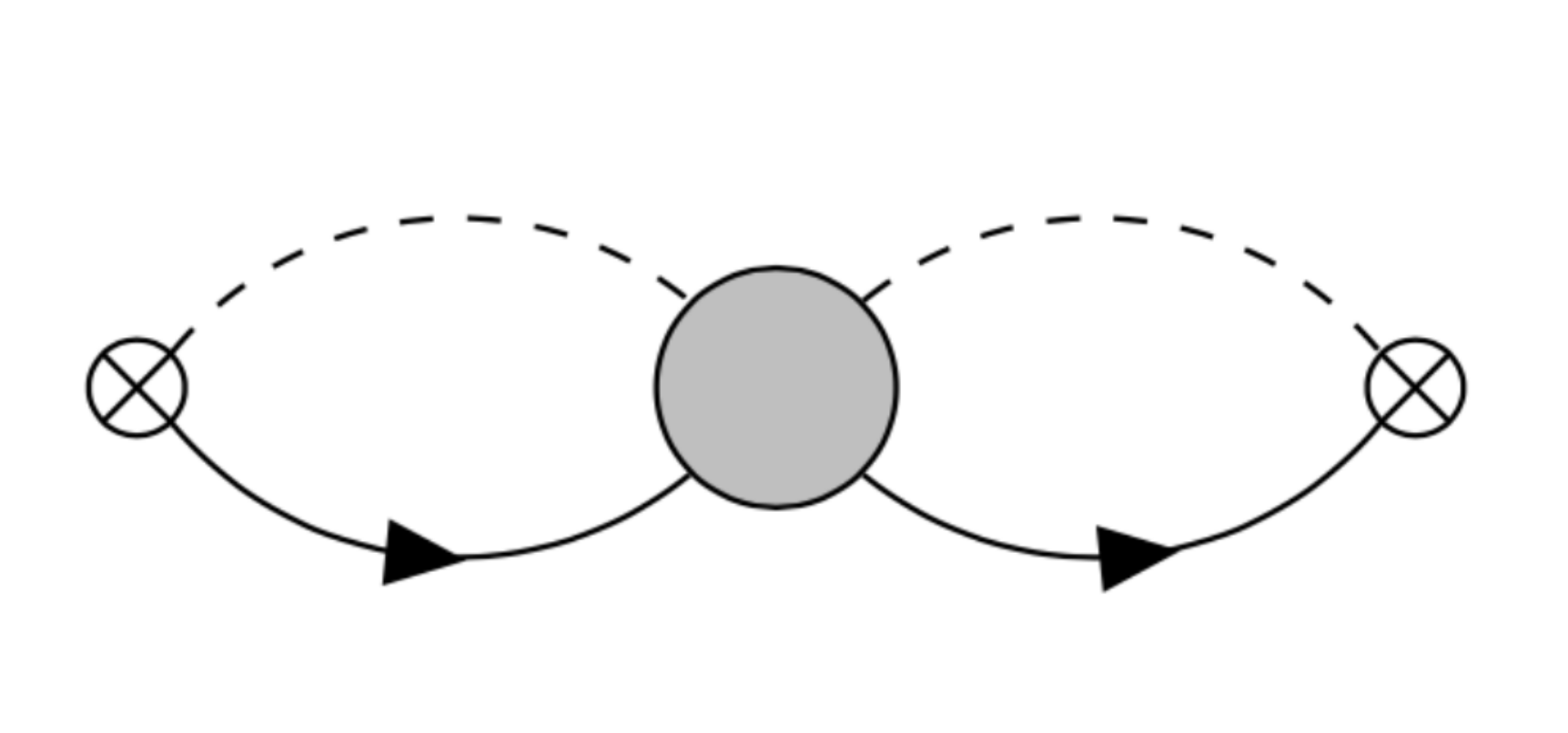}
  \caption{The diagram for the invariant or missing-mass distribution. The crossed dots represent the process which creates (annihilates) the meson-baryon channel.}
  \label{fig:inv_mass}
\end{figure}

In order to extract information on resonances from experimental data, we must link the experimental observables to analytic functions, such as the T-matrix or the Green's function.
Here, we briefly review their relations and explain our notations used in the present paper, having in mind the resonances in the meson-baryon system.
The most typical observable to explore the resonances in the meson-baryon system is the meson-baryon scattering cross section, $\sigma$, as shown in Fig.~\ref{fig:cross}.
$\sigma$ is related to the imaginary part of the T-matrix via the optical theorem as,
\begin{equation}\label{eq:optical}
  \sigma\propto\textrm{Im}\mathcal{T}.
\end{equation}
We can also think of other observables such as the meson-baryon invariant or missing-mass distribution, which are obtained by selecting the meson-baryon channel of interest as a part of the final states of some reaction experiment as shown in Fig.~\ref{fig:inv_mass}. The amplitude of the distribution, $\mathcal{N}$, is proportional to the imaginary part of the Green's function as~\cite{Fetter:1971,Bertsch:1975zz,Morimatsu:1994sx},
\begin{equation}\label{eq:inv_mass}
 \mathcal{N} \propto \textrm{Im}\mathcal{G}.
\end{equation}

Since the full Green's function, $\mathcal{G}$, the free Green's function, $\mathcal{G}_0$, and the $T$-matrix, $T$,  are related with each other by the relation,
\begin{figure}
  \begin{align}\label{eq:green}
      \mathcal{G}
      =
      \raisebox{-0.7cm}{
        \includegraphics[width=0.2\linewidth]{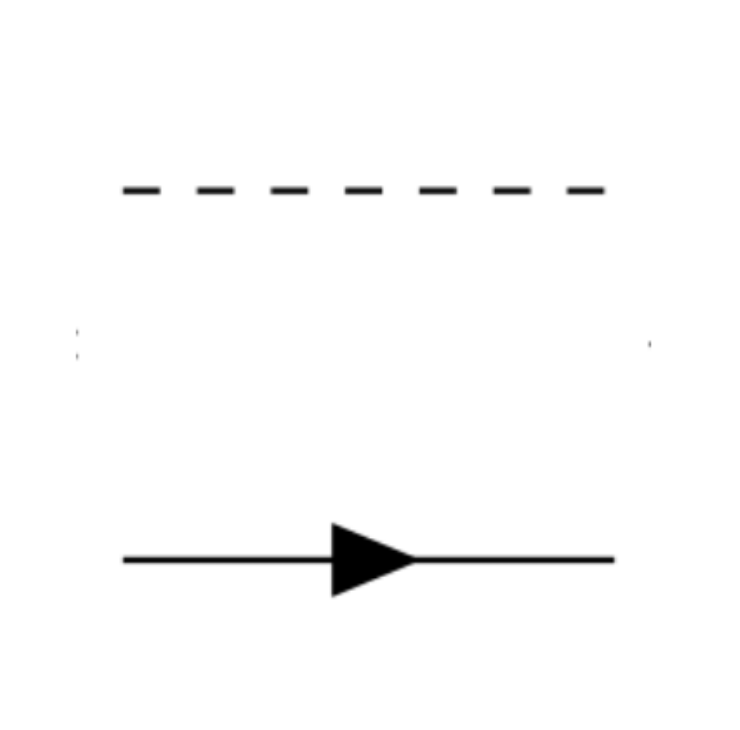}
        }
      +
      \raisebox{-0.7cm}{
        \includegraphics[width=0.2\linewidth]{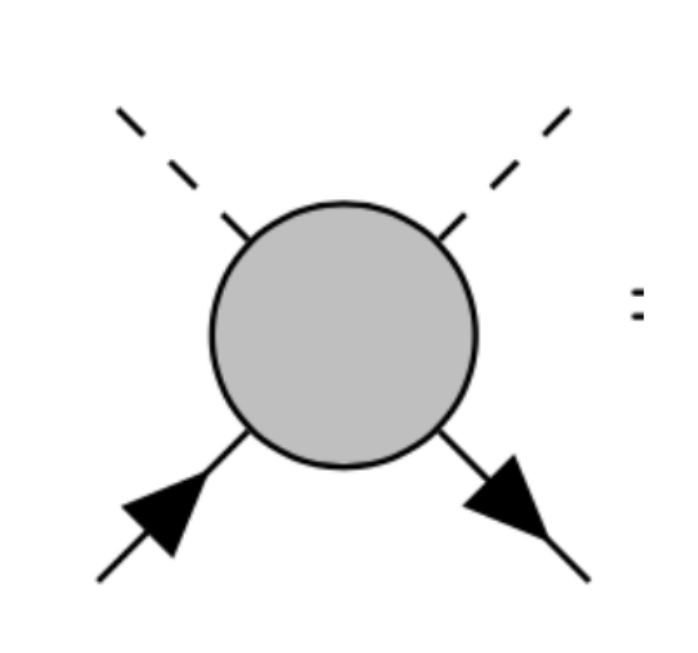}
        }
      =\mathcal{G}_0+\mathcal{G}_0i\mathcal{T}\mathcal{G}_0,
  \end{align}
  \end{figure}
  the full Green's function should have the same analytic structure as the T-matrix. Since our argument is based on the analytic structure of Riemann surfaces, the following argument can be applied to either cases. For our convenience, we will use the Green's function in the following discussion.
  \par
\begin{figure}[h]
      \centering
      \includegraphics[width=0.7\linewidth]{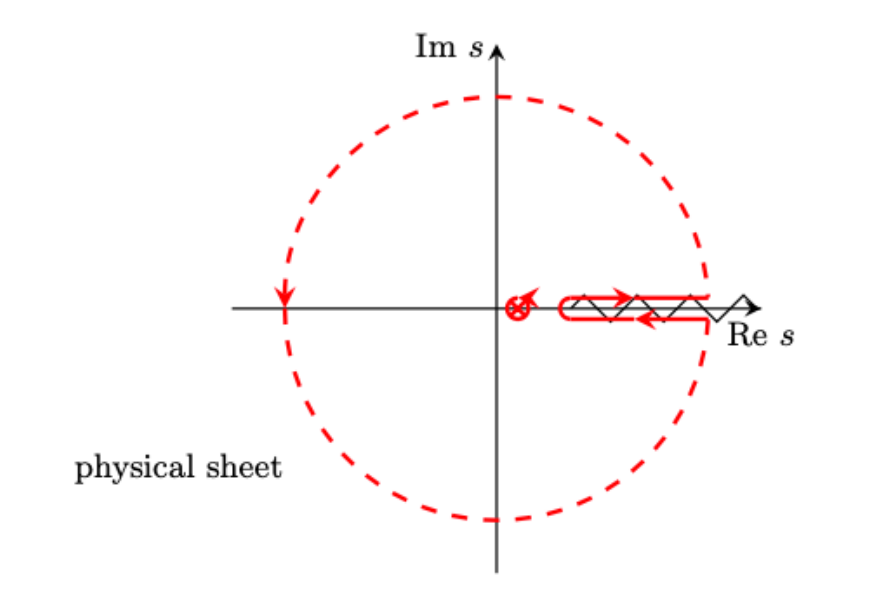}
    \caption{The contour corresponding to the spectral representation Eq.~(\ref{eq:green}). The contour detours the branch cuts that runs along the real axis.}
    \label{fig:spec}
      \end{figure}

We start from the spectral representation of the Green's function. The Green's function can be written as the sum of contributions of the bound states, $\ket{\phi_B} (s=s_B)$, and the continuum, $\ket{\phi_C} (s=s_C)$, as
  \begin{equation}\label{eq:green}
      \mathcal{G}(s)=\sum_{B}\frac{\ket{\phi_B}\bra{\phi_B}}{s-s_B}+\int_{s_{th}}^\infty ds_C \frac{\ket{\phi_C}\bra{\phi_C}}{s-s_C}.
  \end{equation}
  This expansion corresponds to the contour shown in Fig.~\ref{fig:spec}, which detours the branch cuts that run from each threshold to infinity on the complex $s$-plane.
    \par
   In general the Green's function also has resonant poles, which are located on the unphysical sheet of the complex $s$-plane.
   Since the Green's function on the \textit{unphysical} sheets is just an analytic extension of the Green's function on the physical sheet, the contribution from a pole on a \textit{unphysical} sheet cannot be written in the same manner as the contribution from the physical sheet as
    \begin{equation}\label{eq:green'}
        \mathcal{G}(s) \neq \sum_{B}\frac{\ket{\phi_B}\bra{\phi_B}}{s-s_B} + \sum_{R}\frac{\ket{\phi_R}\bra{\tilde\phi_R}}{s-s_R} + \cdots,
    \end{equation}
  where $s_R$ is the pole position on the unphysical sheet of the complex $s$-plane and $\ket{\phi_R}$ and $\ket{\tilde\phi_R}$ are the biorthogonal state vectors~\cite{Sternheim:1972zz}.
    The information about resonant poles, such as complex energies or residues, are only implicitly encoded in the continuum contribution in Eq.~(\ref{eq:green}).
    To decode the information about resonant poles, one must consider a different variable that unfolds the Riemann sheets. this process of unfolding the Riemann surface is called \textit{uniformization}~\cite{Newton}. Uniformization is essential to treat the bound state poles and resonant poles in the same manner.\\
  \par
  Before we discuss uniformization in detail, we will note an important property regarding the pole position and residue of the S-matrix.
  The S-matrix satisfies the following condition,
  \begin{equation}\label{eq:jost}
    \overline{\mathcal{S}}(\{-\overline{k}\})=\mathcal{S}(\{k\}),
  \end{equation}
where $\{k\}$ represents the set of channel momentum, that is $\{k\}=k$ in single-channel systems and $\{k\}=k_1, k_2$ in double-channel systems.
  From Eq. (\ref{eq:jost}), the poles of the S-matrix (and so does the Green's function) are symmetric with respect to imaginary axis.  Now consider that the Green's function has a pole at $\{k_0\}$ with residue $c_0$. The symmetry properties imply that there is a pole at $\{k\}=\{-\overline{k}_0\}$. By keeping in mind of Eq.~(\ref{eq:jost}) and the orientation of a contour around $\{k\}=\{-\overline{k}_0\}$, it can be shown that the residue at $k=\{-\overline{k}_0\}$ is $-\overline{c}_0$.
  To summarize, the poles of the Green's function form a symmetric pair about the imaginary axis and the residues are related by the complex conjugate of its counterparts. \\
  \par
\begin{figure}[h]
  \centering
    \begin{tabular}{c}


      \begin{minipage}{0.50\hsize}
        \centering
        \includegraphics[width=1.0\linewidth]{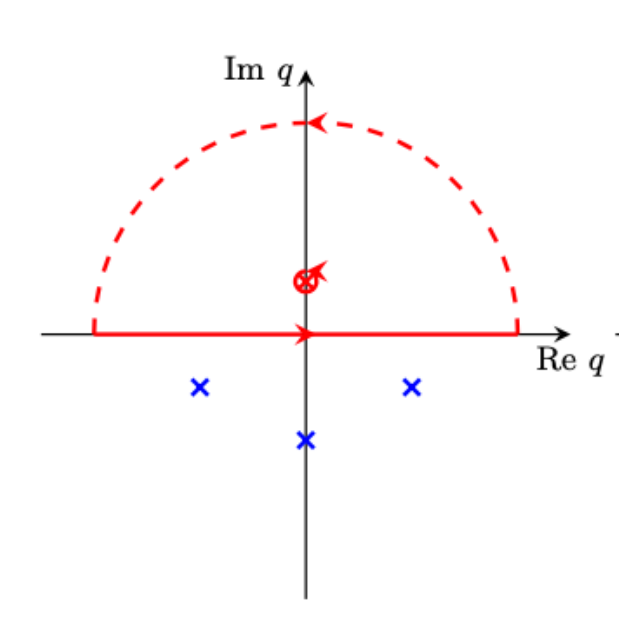}
        \subcaption{}
        \label{fig:q-spec}
      \end{minipage}


      \begin{minipage}{0.50\hsize}
        \centering
        \includegraphics[width=1.0\linewidth]{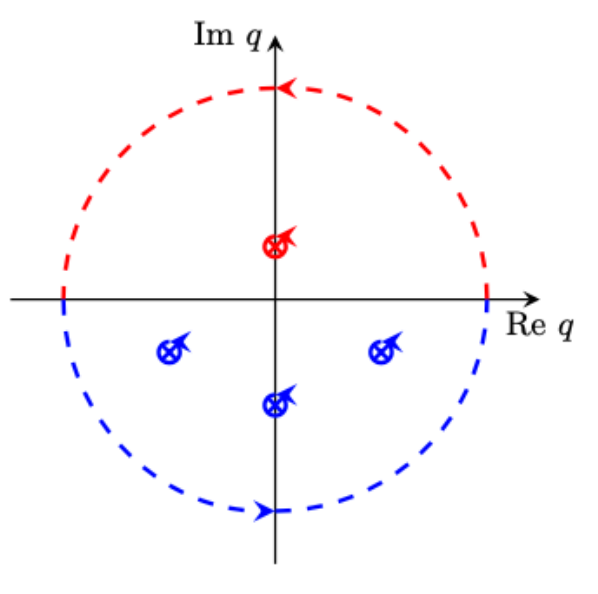}
        \subcaption{}
        \label{fig:q-det}
      \end{minipage}

    \end{tabular}
    \caption{Contours in the $q$-plane. Fig. \ref{fig:q-spec} corresponds to the spectral representation Eq. (\ref{eq:green}), (Fig. \ref{fig:spec}).
    By deforming the contour as Fig. \ref{fig:q-det}, we obtain the pole-sum representation, Eq. (\ref{eq:single}).}
  \end{figure}

  Now let us move on to the details of uniformization. The appropriate kinetic variable to uniformize the Riemann surface depends on the number of channels considered. For single-channel systems, we define a dimensionless variable $q$ by
  \begin{equation}\label{eq:momenta}
    q=\sqrt{s-\varepsilon^2}=\sqrt{\frac{\varepsilon}{\mu}}k+{\mathcal O}(k^3),
  \end{equation}
  where $\varepsilon$ is the threshold energy and $\mu$ is the reduced mass.
  $q$ is proportional to the momentum $k$, at the threshold. The spectral representation (Fig. \ref{fig:spec}) corresponds to the contour in Fig. \ref{fig:q-spec}. To explicitly write down the contributions from the resonant poles, we deform the contour into the the \textit{unphysical} domain shown in Fig. \ref{fig:q-det}.
  \par
  The contour in Fig. \ref{fig:q-det} picks up contributions from all the bound and resonant poles so that the Green's function can be written by a simple sum formation as
  \begin{equation}\label{eq:single}
    \mathcal{G}(q)=\sum_n\frac{\ket{\phi_n}\bra{\tilde\phi_n}}{q-q_n}.
  \end{equation}
   (The pole-sum representation of the Green's function at finite temperature is discussed in Ref.~\cite{Hidaka:2003mm}).\\
  The advantage of this representation is that each component of the series is explicitly written in simple form by the residue and position of the pole. Now by imposing the symmetry condition onto Eq.~(\ref{eq:single}), we can write it in the form of the sum of contributions from each pair as
  \begin{equation}\label{eq:pair}
    \mathcal{G}(q)=\sum_n \mathcal{A}_n(q)=\sum_n\left(\frac{c_n}{q-q_n}-\frac{\overline{c}_n}{q+\overline{q}_n}\right),
  \end{equation}
  where $n$ denotes each pair. We call this the uniformized pole-sum representation.
  One notable point of Eq.~(\ref{eq:pair}) is by imposing the pole symmetry properties, the threshold behavior of each pair contribution naturally behaves in the proper way except for the positive definite property. The positive-definiteness is satisfied by taking the sum of all pairs. Fundamentally, the pole symmetries and the threshold behaviors both originate from the same condition: the unitarity of the S-matrix.\\ Under the threshold, $q$ is purely imaginary. Since $A_n$ can be written by Eq. (\ref{eq:pair2}), the contribution from a pair is identically zero.
  \begin{equation}\label{eq:pair2}
    \textrm{Im} \mathcal{A}_n(q)=\textrm{Im}\frac{c_n(q+\overline{q})}{|q|^2+q_n(q-\overline{q})-q_n^2}.
  \end{equation}
  Right above the threshold, $A_n$ can be written as Eq. (\ref{eq:thr_1}). The pair contributes linearly in terms of $q$.
  \begin{align}
  \textrm{Im}\mathcal{A}_n(q)=
  \begin{cases}
    0, & (\sqrt{s}<\varepsilon) \\
    \displaystyle{-\textrm{Im}\frac{2c_n}{q_n^2}q+\mathcal{O}(q^2)}, &  (\sqrt{s}>\varepsilon)
  \end{cases}
  \label{eq:thr_1}
  \end{align}
  Eq.~(\ref{eq:thr_1}) coincides with Eq.~(\ref{eq:th1}).
  \par
\begin{figure}[h]
      \centering
      \includegraphics[width=0.5\linewidth]{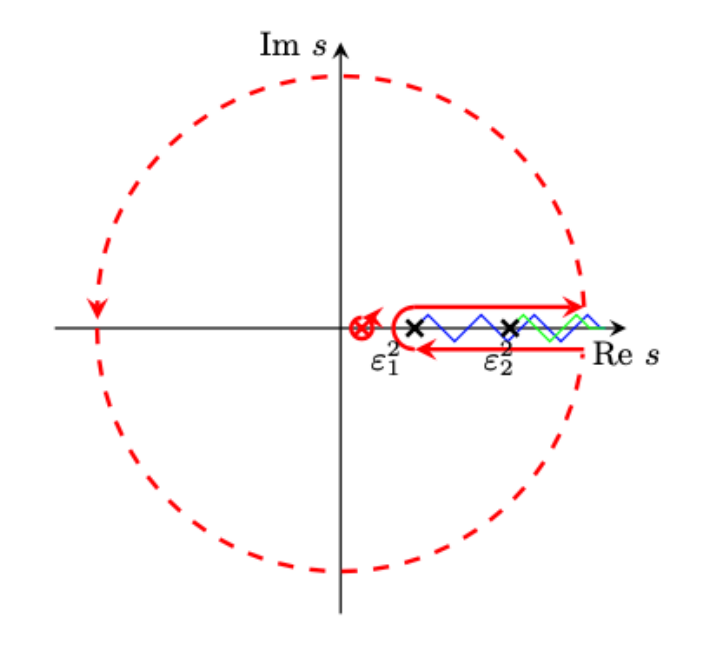}
      \caption{The contour corresponding to the spectral representation, Eq. (\ref{eq:green}), in the case of a double-channel system. The contour detours two branch cuts that run along the real axis from each threshold, $\varepsilon_1^2$ (blue) and $\varepsilon_2^2$ (green) to $\infty$.}
      \label{fig:double-E}
    \end{figure}

    In the case of a double-channel system, the Green's function has two branches with branch points at threshold energies squared, $\varepsilon_1^2$ and $\varepsilon_2^2$, as shown in Fig. \ref{fig:double-E}. To express the Green's function in pole-sum representation, one must choose an appropriate variable and unfold the four Riemann sheets~\cite{Newton}. The basic strategy is as follows. By the change of variables, we send one of the branch points to infinity. Then the structure of the Riemann surface becomes the same as the single-channel case which we can easily unfold.
    \begin{figure}[h]
    \centering
      \begin{tabular}{c}


        \begin{minipage}{0.50\hsize}
          \centering
          \includegraphics[width=1.0\linewidth]{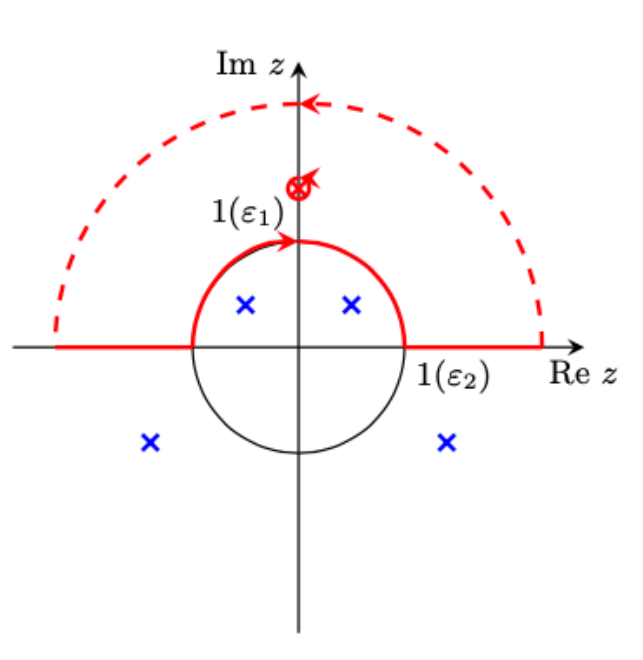}
            \subcaption{}
            \label{fig:double-z}
        \end{minipage}

        \begin{minipage}{0.50\hsize}
          \centering
          \includegraphics[width=1.0\linewidth]{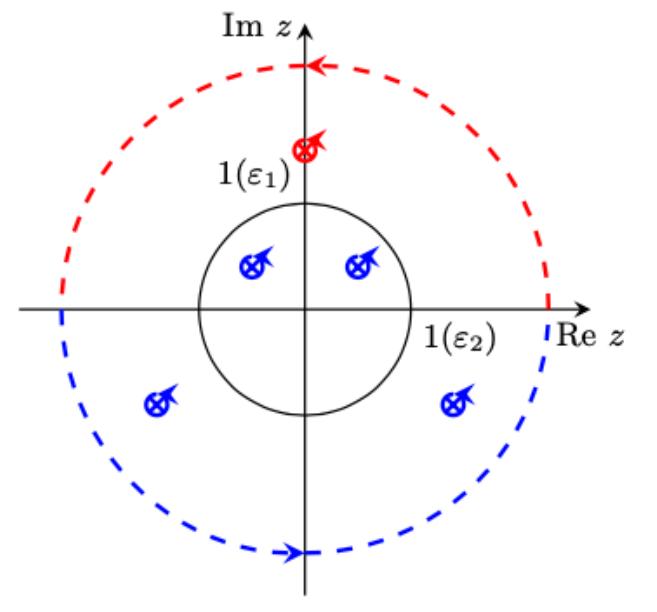}
            \subcaption{}
            \label{fig:double-z_spec}
        \end{minipage}

      \end{tabular}
      \caption{Contours in the $z$-plane. Fig. \ref{fig:double-z} corresponds to the spectral representation Eq. (\ref{eq:green}), (Fig. \ref{fig:spec}).
      By deforming the contour as Fig. \ref{fig:double-z_spec}, we obtain the pole-sum representation.}
      \label{fig:double}
    \end{figure}
    \begin{equation}\label{eq:double}
      z=\frac{1+\sqrt{u}}{1-\sqrt{u}}, \quad u=\frac{q_1-\Delta}{q_1+\Delta},
    \end{equation}
    where, $q_i=\sqrt{s-\varepsilon_i^2}=\sqrt{\frac{\varepsilon_i}{\mu_i}}k_i + \mathcal{O}(k_i^3)$, and $\Delta=\sqrt{\varepsilon_2^2-\varepsilon_1^2}$.
  By the same argument as the single-channel case, the Green's function can be written in pole-sum representation using variable $z$ as
  \begin{equation}\label{eq:double_ps}
    \mathcal{G}(z)=\sum_n  \mathcal{A}_n(z)=\sum_n\left(\frac{c_n}{z-z_n}-\frac{\overline{c}_n}{z+\overline{z}_n}\right).
  \end{equation}
  Note that when $q_1\to-\overline{q}_1$ and $q_2\to-\overline{q}_2$, $z\to-\overline{z}$. Thus, the same symmetric conditions hold for the poles in the $z$-plane:
  \begin{equation}\label{eq:jost''}
    \overline{\mathcal{S}}(-\overline{z})=\mathcal{S}(z).
  \end{equation}
  The threshold behaviors are given in the vicinity of $\sqrt{s}=\varepsilon_1$ as
    \begin{align}
      \textrm{Im} \mathcal{A}_n(z) =
      \begin{cases}
          0, & (\sqrt{s} < \varepsilon_1)\\
          \displaystyle{-\textrm{Im}\frac{2c_n}{(z_n-i)^2}\frac{q_1}{\Delta}+\mathcal{O}(q_1^2)}, & (\sqrt{s} > \varepsilon_1)
      \end{cases}
      \label{eq:thr_10}
  \end{align}
   and in the vicinity of $\sqrt{s}=\varepsilon_2$ as
     \begin{align}
      &\textrm{Im} \mathcal{A}_n(z) = \nonumber\\
      &
      \begin{cases}
          \displaystyle{\textrm{Im}\frac{2c_n}{1-z_n^2}-\textrm{Re}\frac{4c_nz_n}{(1-z_n^2)^2}\frac{\tilde{q_2}}{\Delta}+\mathcal{O}(\tilde{q}_2^2)}, &  (\sqrt{s} < \varepsilon_2) \\
           \displaystyle{\textrm{Im}\frac{2c_n}{1-z_n^2}-\textrm{Im}\frac{2c_n(1+z_n^2)}{(1-z_n^2)^2}\frac{q_2}{\Delta}+\mathcal{O}(q_2^2)}, &  (\sqrt{s} > \varepsilon_2)
      \end{cases}
    \label{eq:thr_20}
    \end{align}
  where $\tilde{q_2}$ is defined by $q_2=i\tilde{q_2}$. Eqs.~(\ref{eq:thr_10}) and (\ref{eq:thr_20}) coincide with Eqs.~(\ref{eq:th1}) and (\ref{eq:th2}), respectively.\\
  \par
  For systems with three or more channels, the Riemann surface of the Green's function cannot be uniformized into a single complex plane. For example, the Riemann surface of a three-channel system is topologically equivalent to a torus. In these cases, there is no simple variable to express the Green's function in the form of Eq.~(\ref{eq:single}). Nevertheless, one can unfold a local region which may be sufficient when considering a particular energy region.\\
  \par
  To summarize, by appropriately uniformizing the Riemann surface, we can expand the Green's function or T-matrix by the uniformized pole-sum representation. The symmetry conditions on the poles naturally lead to the proper threshold behaviors.\\
  \par

  Based on the observation above that observables can be expressed as the imaginary part of the sum of all pole terms in the uniformized complex plane, we propose the following procedures to extract information of the complex energies and residues of the resonance poles from observables in a model-independent manner.
  \begin{enumerate}[i]
    \item Find an appropriate kinetic variable, $z$, that uniformizes the system.
    \item Assume that the amplitude $\mathcal{A}(z)$, whose imaginary part gives observables such as the cross section $\sigma$, or the missing-mass distribution $d\sigma/dm$, is approximated by a few ($m$) pairs of the pole terms as
    \begin{equation}\label{eq:sumform}
      \mathcal{A}(z)=\sum_{n=1}^m \left(\frac{c_n}{z-z_n} - \frac{\overline{c}_n}{z-\overline{z}_n}\right).
    \end{equation}
    \item Determine the complex positions and residues, $z_n$ and $c_n$, ($n=1,\cdots, m$), by fitting $\textrm{Im} \mathcal{A}(z)$ to the experimental data, from which the complex energies and residues of resonances are obtained.
  \end{enumerate}

  Here, we clarify the reliability of the obtained results.
  If one increases the number of pairs of poles in the sum, the position of the complex poles and residues, $z_n$ and $c_n$, would change in general.
  If the fitting is successful, then $z_n$ and $c_n$ would change little for the poles near the fitting energy region, while they might change to some extent for the poles away from the fitting energy region.
   Besides, the position of the newly added pairs should be most away from the fitting energy region.
Ideally, if this condition is met, one could conclude that the fitting is successful, and the obtained complex positions and residues for the poles near the fitting energy region are regarded as reliable results.
  \par
  The counterpart of the resonance pole may be very apart from the physical region of interest. Then, its contribution is negligible, and the contribution of the resonance pole alone gives the threshold behavior in practice. Nevertheless, the inclusion of the counterpart of the resonance pole does not increase the number of parameters and therefore does nothing wrong.
  \par
  Also, note that the method is model-independent in the sense that the procedure does not depend on a particular model.
  The only necessary information is the channels of relevant particles.\\

  The eventual goal of this project is to extract information of near-threshold resonances from experimental data by applying the method just explained above.
  In this paper, however, we apply our method to the results of model theory.
  Since we can precisely calculate the pole energies and residues within the model theory, we can test if our method works or not by comparing the fitted results with the exact ones.

  \par
\begin{figure}[h]
  \centering
    \includegraphics[width=0.7\linewidth]{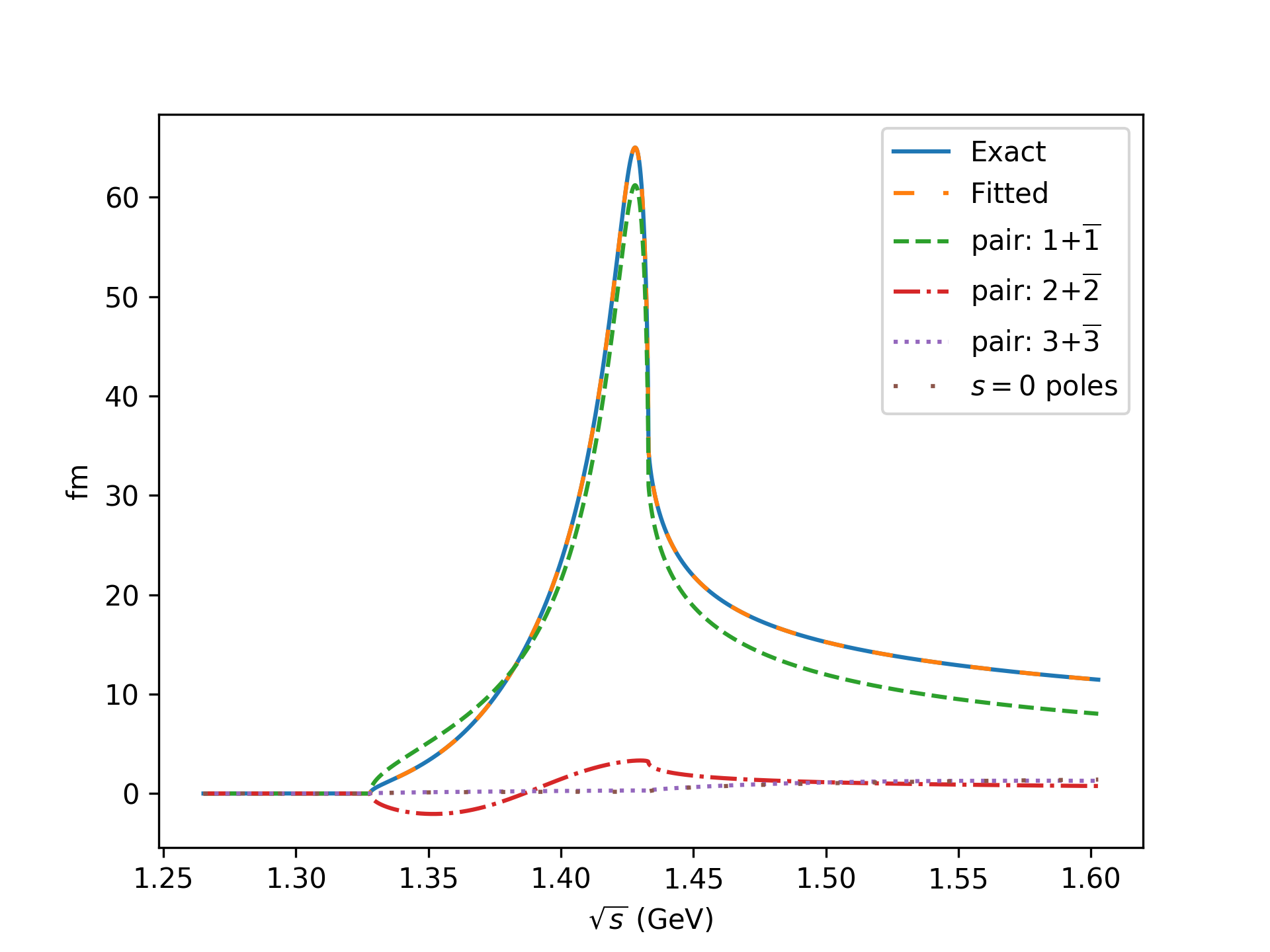}
    \caption{Exact and fitted results of Im$\mathcal{T}$ for the $\overline{K}N-\overline{K}N$ component.
      We also show the contributions from each pair of poles $1+\overline{1}$ , $2+\overline{2}$, $3+\overline{3}$ and the $s = 0$ poles for the fitted Im$\mathcal{T}$.
      The specifications are given in the figure.}
    \label{fig:res__1}
  \end{figure}
\begin{figure}[h]
  \centering
    \includegraphics[width=0.7\linewidth]{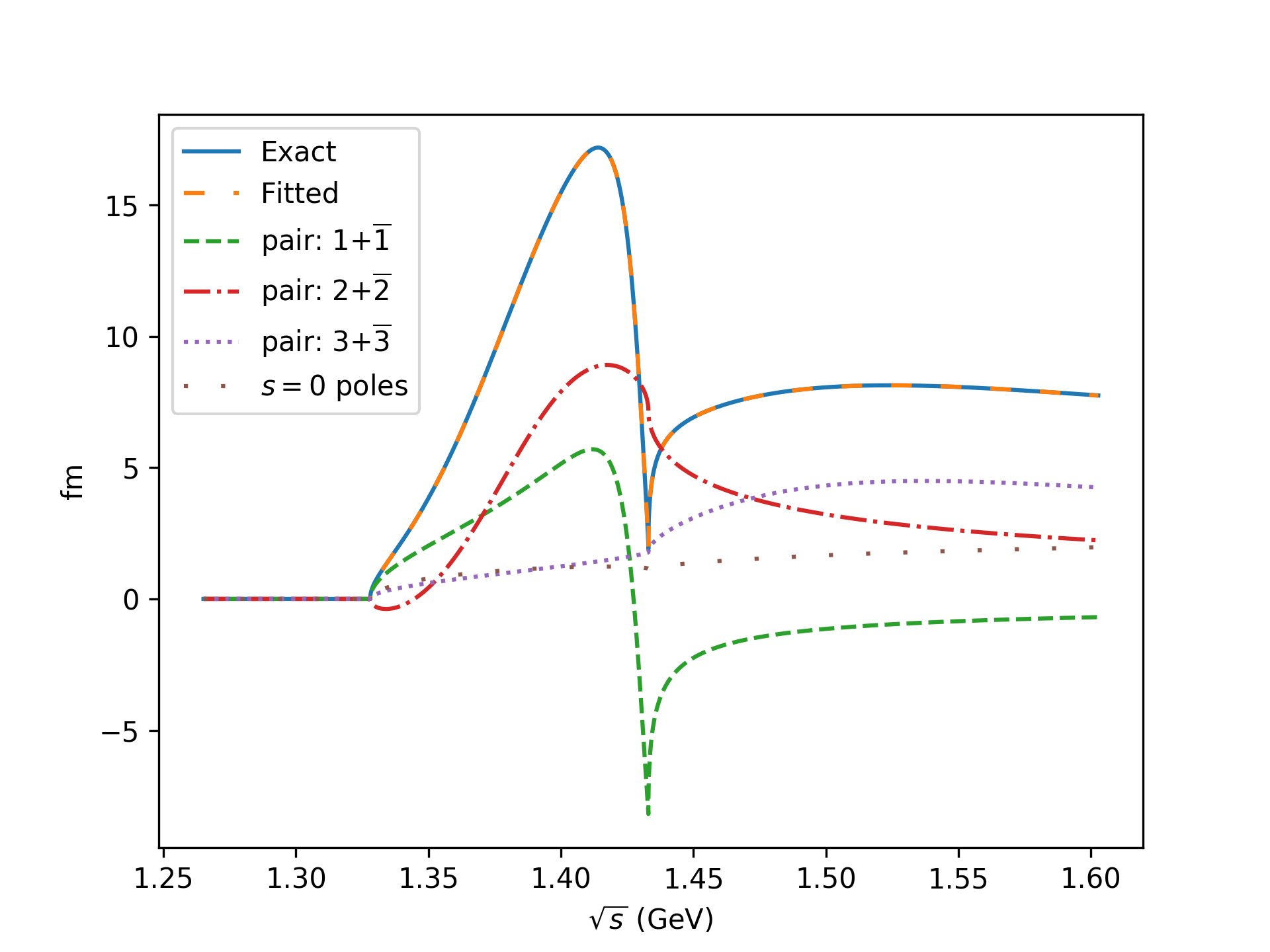}
    \caption{The same as Fig. \ref{fig:res__1} for the $\pi\Sigma-\pi\Sigma$ component.}
    \label{fig:res__3}
  \end{figure}

  For the model theory, we consider a meson-baryon scattering in the chiral-unitary model~\cite{Kaiser:1995eg,Oset:1997it,Hyodo:2011ur,Morimatsu:2019wvk} involving two channels, $\overline{K}N(I=0)$ and $\pi\Sigma(I=0)$. Details about the numerical calculation will be shown elsewhere~Ref.~\cite{Morimatsu:to_be_published}.
  The imaginary part of the calculated scattering $T$-matrix, Im$\mathcal{T}$, is shown in Figs.~\ref{fig:res__1}-\ref{fig:res__3}, which we regard as virtual experimental data and apply our method.

  \begin{figure}[h]
  \centering
    \includegraphics[width=0.7\linewidth]{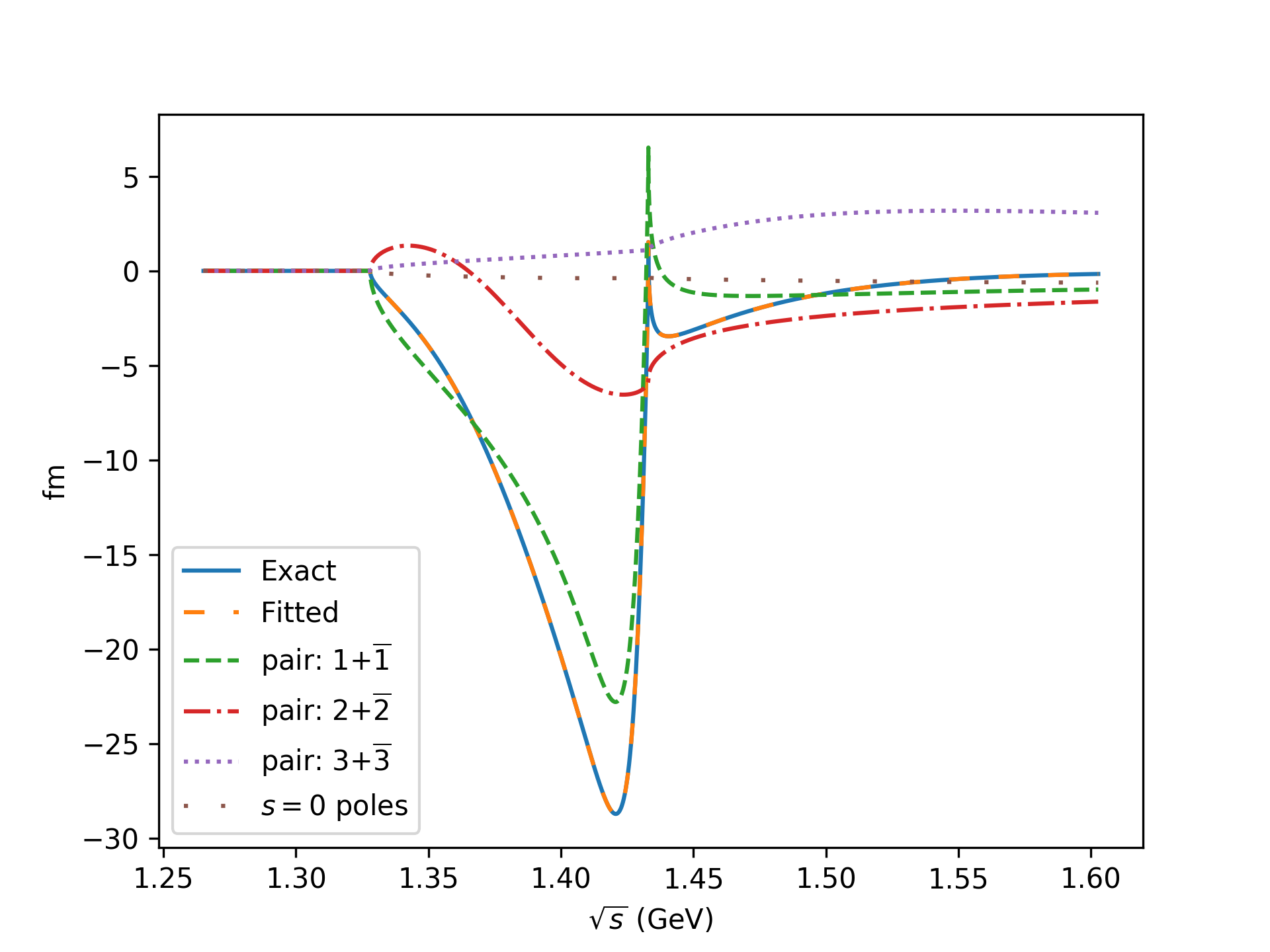}
    \caption{The same as Fig. \ref{fig:res__1} for the $\pi\Sigma-\overline{K}N$ component.}
  \end{figure}
  \begin{figure}[h]
  \centering
    \begin{tabular}{c}
      \begin{minipage}{0.50\hsize}
        \centering
        \includegraphics[width=\linewidth]{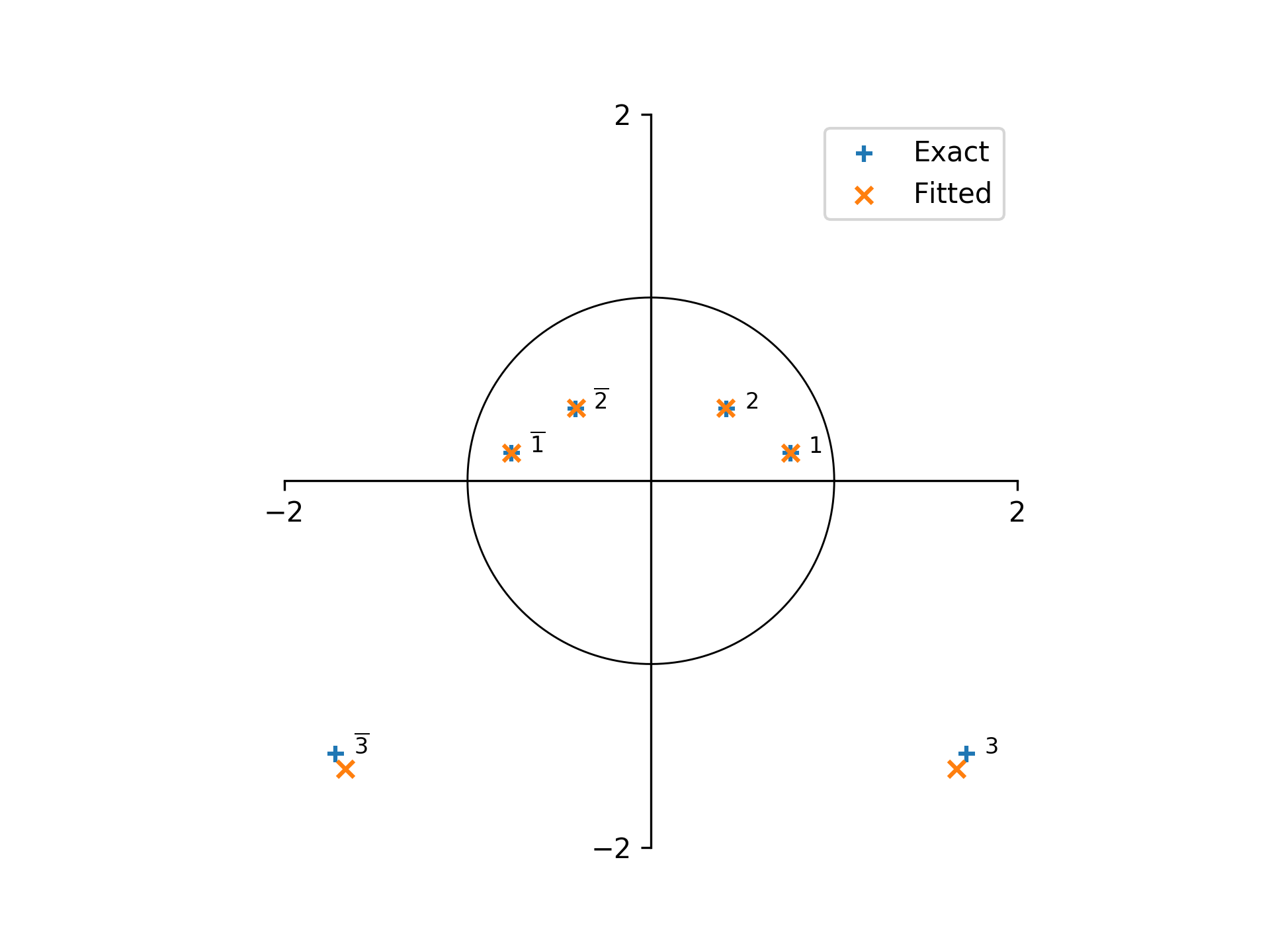}
      \end{minipage}

      \begin{minipage}{0.50\hsize}
        \centering
        \includegraphics[width=\linewidth]{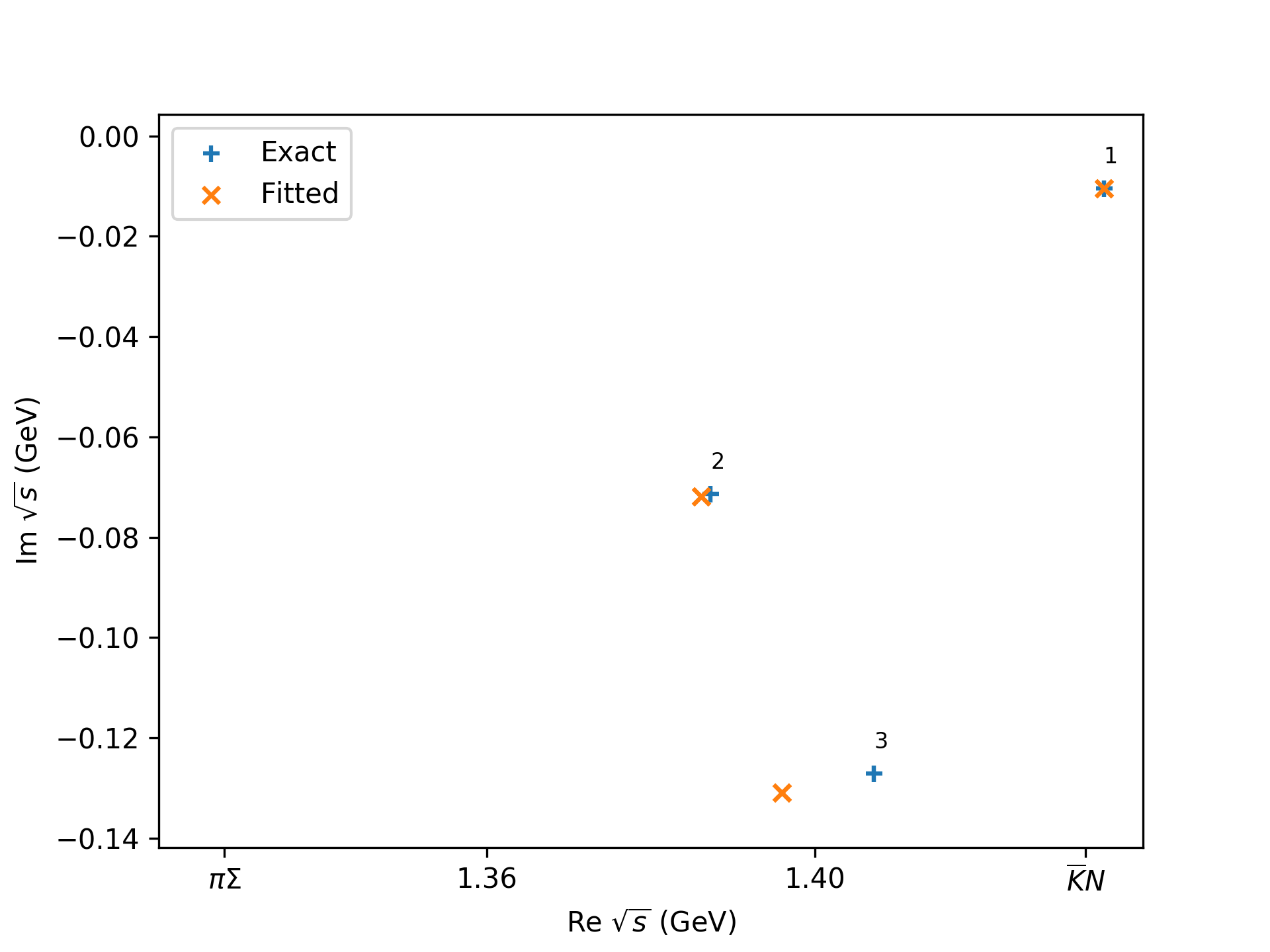}
      \end{minipage}
    \end{tabular}
    \caption{Exact ($+$) and fitted ($\times$) results of positions of the poles in the $z$-plane(left) and $\sqrt{s}$-plane(right). Three poles ($1$, $2$, $3$) in the neighborhood of the thresholds are shown in both planes and their symmetric counterparts ($\overline{1}$, $\overline{2}$, $\overline{3}$) are also shown in the $z$-plane.}
    \label{fig:po}
  \end{figure}
  \begin{table*}
    \begin{tabular}{cccccc}
      &&pole position($z$)& $\pi\Sigma-\pi\Sigma$ &$\pi\Sigma-\overline{K}N$ &$\overline{K}N-\overline{K}N$ \\ \hline
      \multirow{2}{*}{pole: $1$, $\bar1$}&& $\pm$0.760+0.154i&-0.901$\pm$2.72i&-1.56$\mp$6.52i&11.4$\pm$4.79i\\
      &&($\pm$0.760+0.154i)&(-0.892$\pm$2.72i)&(-1.57$\mp$6.51i)&(11.4$\pm$4.78i)\\ \\
      \multirow{2}{*}{pole: $2$, $\bar2$}&&$\pm$0.409+0.397i&3.27$\mp$1.99i&-2.35$\pm$2.41i&1.05$\mp$2.10i\\
      &&($\pm$0.413+0.395i)&(3.31$\mp$1.85i)&(-2.41$\pm$2.34i)&(1.11$\mp$2.06i)\\\\
      \multirow{2}{*}{pole: $3$, $\bar3$}&&$\pm$1.67-1.57i&6.00$\pm$3.94i&4.67$\pm$2.31i&2.25$\pm$0.461i \\
      &&($\pm$1.72-1.49i)&(4.83$\pm$3.92i)&(3.70$\pm$2.30i)&(1.95$\pm$0.814i) \\\hline
    \end{tabular}
    \caption{Fitted poles and residues of $i\mathcal{T}$ in the $z$-plane. the numbers in the
    parentheses are the numerically calculated results.}
    \label{tab:res2}
  \end{table*}
  \begin{table*}
    \begin{tabular}{cccccc}
      &&pole position ($\sqrt{s}$: GeV)& $\pi\Sigma-\pi\Sigma$ &$\pi\Sigma-\overline{K}N$ &$\overline{K}N-\overline{K}N$ \\ \hline
      \multirow{2}{*}{pole: $1$, $\bar1$}&&1.435$\mp$0.010i&$\pm$31.7-10.7i&$\mp$49.9+60.2i&$\mp$35.2-140.i\\
      &&(1.435$\mp$0.010i)&($\pm$31.6-10.7i)&($\mp$49.7+60.2i)&($\mp$35.3-140.i)\\ \\
      \multirow{2}{*}{pole: $2$, $\bar2$}&&1.386$\mp$0.072i&$\pm$1.81-12.1i&$\pm$1.12+10.7i&$\mp$3.02-6.90i\\
      &&(1.387$\mp$0.071i)&($\pm$2.03-12.1i)&($\pm$1.03+10.8i)&($\mp$2.95-6.907i)\\\\
      \multirow{2}{*}{pole: $3$, $\bar3$}&&1.397$\mp$0.131i&$\pm$18.7+55.4i&$\pm$18.3+38.3i&$\pm$12.1+14.3i\\
      &&(1.407$\mp$0.127i)&($\pm$13.0+49.8i)&($\pm$13.4+33.5i)&($\pm$9.01+15.0i) \\\hline
    \end{tabular}
    \caption{Fitted poles and residues of $i\mathcal{T}$ in the $\sqrt{s}$-plane. the numbers in the
    parentheses are the numerically calculated results.}
    \label{tab:res3}
  \end{table*}
  \par
  Since the system is a double-channel system, the appropriate kinetic variable is given by $z$ in Eq. (\ref{eq:double}).
  In addition to the resonant poles, there is a pole at $s=0$ (corresponding to two poles $z_{\pm}$ on the $z$-plane) originating from relativistic kinematics.
  \begin{gather}
    z_\pm=\frac{\sqrt{\pm i\varepsilon_1+\Delta}+\sqrt{\pm i\varepsilon_1-\Delta}}{\sqrt{\pm i\varepsilon_1+\Delta}-\sqrt{\pm i\varepsilon_1-\Delta}} \label{eq:s_0},
  \end{gather}
  where, $\varepsilon_1=m_\pi + m_\Sigma$, $\varepsilon_2=m_{\overline{K}} + m_N$, and $\Delta=\sqrt{\varepsilon_2^2-\varepsilon_1^2}$.
  \par
  In the model we are considering, we already know that there are only three resonant pairs in the neighborhood of the thresholds from the results of exact calculation. Therefore we fitted the T-matrix with three resonant pole pairs and the poles at $z_\pm$. Considering the symmetry of the paired poles, the fitting function can be expressed as
  \begin{equation}\label{eq:fit}
    \mathcal{A}(z)=\sum_{n=1}^3 \left(\frac{c_n}{z-z_n}-\frac{\overline{c}_n}{z+\overline{z}_n}\right)+\sum_{\pm} \frac{d_\pm}{z-z_\pm},
  \end{equation}
  where the fitting parameters are $c_n$, $z_n$, and $d_\pm$ ($d_\pm$ is purely imaginary).
  We have 4 real parameters for each pair of resonance pole: 2 for the pole position and 2 for the residue, and 1 for the residue of each kinematical pole at $z_{\pm}$.
  Thus, we have 14 parameters altogether.\\
  \par
  In Figs.~\ref{fig:res__1}-\ref{fig:res__3}, we plot and compare the results of the exactly calculated Im$\mathcal{T}$, to the fitted  uniformized pole-sum representation.
 We can hardly regard the difference between the fitted results and the exact model calculation.
 Also shown in Fig.~\ref{fig:res__1}-\ref{fig:res__3} are the contributions from each pair of poles and the $s=0$ poles.
 The contribution of the $1+\bar 1$ pair explains most of the $\overline{K}N-\overline{K}N$ component.
 For the $\pi\Sigma-\overline{K}N$ component, the contribution of $1+\bar 1$ pair is still the largest but that of $2+\bar 2$ is also considerable.
 For the $\pi\Sigma-\pi\Sigma$ component, both contributions of $1+\bar 1$ and $2+\bar 2$ are important but the latter is larger than the former.
 All $1+\bar 1$, $2+\bar 2$ and $3+\bar 3$ pairs contribute above the $\overline{K}N$ threshold for the $\pi\Sigma-\overline{K}N$ and $\pi\Sigma-\pi\Sigma$ components.
 The contribution of $s = 0$ poles is insignificant everywhere in the range of interest.
 Thus, even if we do not take into account the contribution from the $s=0$ poles, we expect that the fitted results should hardly change.
 From the results above, we can conclude that the poles $1+\bar 1$ and $2+\bar 2$ explain peak structures and $3+\bar 3$ gives background contributions of the \textit{virtual} experimental data.
\par
 In table \ref{tab:res2} and \ref{tab:res3}, we show the fitted results of the pole positions and residues of the $T$-matrix in the $z$-plane and $\sqrt{s}$-plane respectively. The exactly calculated results in the model theory are also shown for comparison for poles 1, 2, and 3.
 Also, the positions of poles are mapped in the complex $z$-plane and in the complex $\sqrt{s}$-plane, respectively, in Fig. \ref{fig:po}.
 From these results, we can acknowledge that the fitted positions and residues of poles agree very well with those of the exact calculation.
 Pole $1$ ($\bar 1$) is located close to the real $\sqrt{s}$ axis, and the difference between the fitted and exact results is minimal.
 As the distance from the real $\sqrt{s}$ axis increases, for pole $2$ ($\bar 2$) and further for pole $3$ ($\bar 3$), the difference between the fitted and exact results widen.
 However, even for pole $3$ ($\bar 3$), the difference is still rather small.
 We also notice that the difference seems to enlarge when moving from the complex $z$-plane to the complex $\sqrt{s}$-plane.\\
 \par
 To summarize, the imaginary part of the T-matrix can be expressed extremely well with the uniformized pole-representation. Also, the fitted poles and residues are in good agreement with the numerically exact result.
 \par
 From these results, we conclude that if experimentalists could provide us with data of sufficient quality and quantity, we can perfectly reproduce well the experimental data, and determine the complex energies and the residues in a completely model-independent manner. The method accurately reproduces not only the peak structures but also continuous spectra with proper threshold behaviors, in a natural way. Realistically, however, the experimental situation may not be perfect. Even in such a situation, the use of uniformized pole-sum representation would provide us with a framework which is theoretically more reasonable and practically more useful than usual methods in the sense that it automatically incorporates the proper threshold behaviors.\\
 \par
 The authors would like to thank Kazuki Yamada for the discussion at the early stage of the work.
 Osamu Morimatsu would also like to thank the members of the discussion meeting held in KEK Tokai campus, Yoshinori Akaishi, Akinobu Dote, Toru Harada, Fuminori Sakuma, and Shoji Shinmura.

  \end{document}